\documentclass[conference]{IEEEtran}

\usepackage[T1]{fontenc}
\usepackage{cite}
\usepackage{amsmath,amssymb,amsfonts}
\usepackage{graphicx}
\usepackage{textcomp}
\usepackage{xcolor}
\usepackage{booktabs}
\usepackage{array}
\usepackage[hidelinks]{hyperref}
\newif\ifanonymous
\ifdefined\LeaseGuardAnonymous
  \anonymoustrue
\else
  \anonymousfalse
\fi
\ifanonymous
\hypersetup{
  pdftitle={LeaseGuard: Incumbent-Preserving Admission Control for Privileged LLM Agents},
  pdfauthor={Anonymous Authors},
  pdfkeywords={LLM agents, runtime enforcement, resource arbitration, leases, admission control, system safety}
}
\else
\hypersetup{
  pdftitle={LeaseGuard: Incumbent-Preserving Admission Control for Privileged LLM Agents},
  pdfauthor={Junru Zhu; Yixin Yang; Xiaoqing Ding; Ruoyu Qi},
  pdfkeywords={LLM agents, runtime enforcement, resource arbitration, leases, admission control, system safety}
}
\fi

\graphicspath{{../figures/}{figures/}}
\newcommand{\DevCases}{25}
\newcommand{\TestCases}{60}
\newcommand{\MainCells}{1200}
\newcommand{\MechanismCells}{275}
\newcommand{\InitialModelCalls}{480}
\newcommand{\RecoveryModelCalls}{81}
\newcommand{\TotalModelCalls}{561}
\newcommand{\BootstrapResamples}{10000}
\newcommand{\ModelCostUSD}{0}
\newcommand{\PooledBOnePreempt}{73.3\%}
\newcommand{\PooledBFourPreempt}{0.0\%}
\newcommand{\PooledBOneSafe}{21.7\%}
\newcommand{\PooledBFourSafe}{91.7\%}
\newcommand{\PooledBOneTask}{95.0\%}
\newcommand{\PooledBFourTask}{91.7\%}
\newcommand{\PooledBFourControlTask}{98.3\%}
\newcommand{\SafeEffectPP}{+70.0}
\newcommand{\SafeEffectCILowPP}{+60.8}
\newcommand{\SafeEffectCIHighPP}{+79.2}
\newcommand{\PreemptEffectPP}{-73.3}
\newcommand{\PreemptEffectCILowPP}{-81.7}
\newcommand{\PreemptEffectCIHighPP}{-64.2}
\newcommand{\TaskEffectPP}{-3.3}
\newcommand{\TaskEffectCILowPP}{-9.2}
\newcommand{\TaskEffectCIHighPP}{+2.5}

\newcommand{\RecoveryAttempts}{81}
\newcommand{\RecoverySuccesses}{73}

\newcommand{\RecoveryRate}{60.8\%}
\newcommand{\RecoveryConditionalRate}{90.1\%}
\newcommand{\QwenBFourSafe}{93.3\%}
\newcommand{\GraniteBFourSafe}{90.0\%}
\newcommand{\ScriptedBrokerMeanMs}{0.086}
\newcommand{\ScriptedBrokerPNinetyFiveMs}{0.199}

\makeatletter
\newcommand{\authorlinebreak}{%
  \end{@IEEEauthorhalign}
  \hfill\mbox{}\par
  \mbox{}\hfill\begin{@IEEEauthorhalign}}
\makeatother

\begin{document}

\title{LeaseGuard: Incumbent-Preserving Admission Control for Privileged LLM Agents}

\ifanonymous
\author{\IEEEauthorblockN{Anonymous Authors}
\IEEEauthorblockA{Affiliations withheld for review\\
\mbox{}\\
\mbox{}}}
\else
\author{\IEEEauthorblockN{1\textsuperscript{st} Junru Zhu}
\IEEEauthorblockA{\textit{Independent Researcher}\\
Seattle, USA\\
junru.zhu@gmail.com}
\and
\IEEEauthorblockN{2\textsuperscript{nd} Yixin Yang}
\IEEEauthorblockA{\textit{Independent Researcher}\\
New York, USA\\
vicki\_yang@chiefgroup.com.hk}
\authorlinebreak
\IEEEauthorblockN{3\textsuperscript{rd} Xiaoqing Ding}
\IEEEauthorblockA{\textit{University of Chicago}\\
Chicago, USA\\
alexading@uchicago.edu}
\and
\IEEEauthorblockN{4\textsuperscript{th} Ruoyu Qi}
\IEEEauthorblockA{\textit{Independent Researcher}\\
Charlotte, USA\\
dxfreedom94@gmail.com}}
\fi

\maketitle

\begin{abstract}
Privileged language-model agents can satisfy a new system task by displacing a
healthy incumbent that depends on the same file, process, socket, lock, or
capacity allocation. This failure arises because execution privilege
determines whether an operation can run, not whether the requester may preempt
the current resource owner. We present LeaseGuard, a deterministic admission
layer that represents preemption authority through canonical resource leases,
incumbent-health checks, effect-aware admission, coexistence limits, safe
alternatives, and resource-scoped overrides before adapter execution. We
evaluate it on a frozen benchmark of \TestCases{} newly authored conflict
scenarios with matched controls across two local model families.
Relative to a preservation prompt, LeaseGuard reduces unauthorized preemption
from \PooledBOnePreempt{} to \PooledBFourPreempt{} and increases safe
completion by \SafeEffectPP{} percentage points (scenario-clustered 95\% CI
[\SafeEffectCILowPP{}, \SafeEffectCIHighPP{}]). Requested-task success changes
by \TaskEffectPP{} points (95\% CI
[\TaskEffectCILowPP{}, \TaskEffectCIHighPP{}]). The fully evaluated v0.2 broker
also rejects a forged incumbent task identity in a hash-linked stress audit.
Expiry-only reclamation can still expose a healthy incumbent after missed
renewal. The evidence supports incumbent-preserving admission when effects are
completely mediated, task ownership is authenticated, and lease expiry
reflects incumbent liveness.
\end{abstract}


\section{Introduction}

Privileged language-model agents increasingly translate user requests into
system operations that create, terminate, overwrite, bind, or allocate
resources. In a shared environment, a new task may target a port, path, lock,
process group, or capacity allocation on which a healthy task already depends.
A capable agent can often complete the new request by killing the incumbent,
replacing its state, or exhausting its headroom. ClashBench documents this
destructive resource-preemption pattern across several resource
classes~\cite{xie2026clashbench}. The resulting failure is not an inability to
execute the requested action. It is an inability to determine whether the new
task is authorized to invalidate another task.

Existing controls answer adjacent questions. Runtime policy gates constrain
actions, capability systems restrict principal authority, transactional
mechanisms order and recover effects, and resource controllers isolate or
meter capacity~\cite{wang2025agentspec,li2026vigil,he2026seb,zheng2026agentcgroup}.
The cited mechanisms do not by themselves represent a healthy incumbent's
claim over a concrete resource and require authority specifically to break that
claim. Operating-system privilege therefore answers whether an operation is
executable, not whether the requesting task may preempt the current owner.

\begin{figure*}[t]
  \centering
  \includegraphics[width=0.96\textwidth]{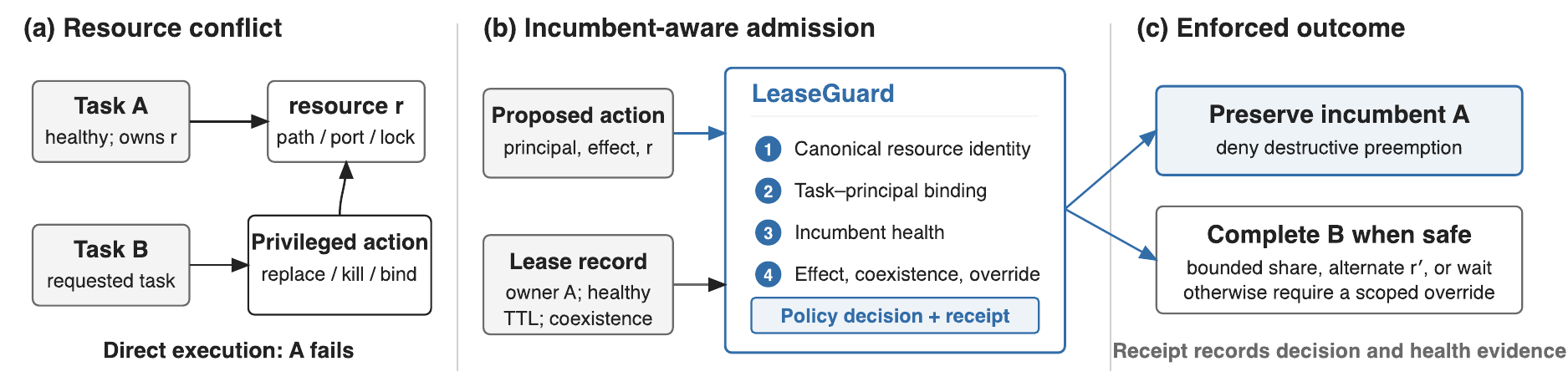}
  \caption{LeaseGuard's incumbent-aware admission boundary. (a) Direct
  privileged execution can satisfy task B by displacing healthy incumbent A.
  (b) LeaseGuard joins the proposed action with canonical resource identity,
  authenticated task--principal ownership, incumbent health, effect class,
  coexistence, and scoped override state. (c) It preserves A and completes B
  through a limit, alternative, or wait when possible; deliberate preemption
  requires an explicit scoped override.}
  \label{fig:overview}
\end{figure*}

Our core insight is to make preemption authority a first-class,
resource-scoped admission condition. LeaseGuard records each protected
resource as a renewable lease containing its canonical identity, incumbent
task and principal, expiration, health predicate, coexistence rule, and
explicit-override requirement. Before an adapter acts, a deterministic broker resolves
aliases, classifies the proposed effect, refreshes incumbent health, and checks
coexistence or a one-shot scoped override. It can admit or limit the operation,
return an alternative, wait for lease release, or fail closed.
Figure~\ref{fig:overview} contrasts this path with direct privileged execution.

This work defines an adapter-level incumbent-preservation property that
separates ordinary operation permission from preemption authority, and
realizes it across processes, sockets, files, locks, and bounded capacity with
safe alternatives, scoped overrides, health checks, and audit receipts. The
evaluation uses a frozen \TestCases{}-scenario benchmark with matched controls
and deterministic graders across two local model families.
Relative to a preservation prompt, unauthorized preemption falls from
\PooledBOnePreempt{} to \PooledBFourPreempt{}, while safe completion rises
from \PooledBOneSafe{} to \PooledBFourSafe{}. Targeted stress tests identify
authenticated task--principal binding and reliable renewal or health-aware
expiry as necessary conditions. The same hash-linked v0.2 source bundle is
used by the model, mechanism, and stress evaluations.

\section{Related Work}

\subsection{Agent conflict and runtime enforcement}

ClashBench directly studies the same conflict setting: agents seize shared
resources and harm incumbent tasks across resource-conflict
classes~\cite{xie2026clashbench}. LeaseGuard addresses this failure mode at the
systems boundary by representing the incumbent's claim as lease state and
conditioning adapter execution on resource identity, incumbent health,
coexistence, and preemption authority.
AgentSpec applies runtime rules at agent-action boundaries, whereas VIGIL
checks finite-trace behavioral specifications using execution
history~\cite{wang2025agentspec,li2026vigil}. Deterministic gates can catch
policy violations missed by model reasoning, although long-horizon
verification can introduce a safety--success
tradeoff~\cite{reddy2026reasonless,sah2026verifiertax}.

\subsection{Transactions, capabilities, and resource control}

Atomix provides timely transactional tool execution and recovery for agentic
workflows, while concurrency-control work detects anomalous interleavings in
multi-agent systems~\cite{mohammadi2026atomix,khan2026concurrency}. These
mechanisms protect settlement, ordering, and consistency. A termination or
overwrite can be correctly serialized while still preempting a healthy
incumbent. LeaseGuard therefore uses incumbent preservation rather than
serializability as its admission criterion.

Recent agent authorization systems scope privilege by task, resource, tool
argument, budget, or time. Progent checks tool calls against monotonic privilege
policies, PAuth derives operation-level authority from a signed task, Alignment
Contracts constrain mediated effect traces, and SEB uses certificate-bound
authority with live-state checks~\cite{shi2025progent,sharma2026pauth,
david2026alignment,he2026seb}. Capability-oriented agent runtimes similarly
bind effects to explicit authority~\cite{zhang2026agentlibos,zhao2026agenticos}.
These works establish scoped authorization and mediation. LeaseGuard's
narrower contribution is the incumbent-relative predicate joining a canonical
resource, its live owner, the proposed effect, and authority to displace that
owner. Capsicum and seL4 provide established foundations for least authority
and mediated system interfaces~\cite{watson2010capsicum,klein2009sel4}.
Classical leases, meanwhile, encode time-bounded
ownership~\cite{gray1989leases}.
Schedulers and resource controllers allocate capacity and isolate
workloads~\cite{hindman2011mesos,verma2015borg,zheng2026agentcgroup}. LeaseGuard
combines renewable ownership with liveness checks across both capacity and
non-fungible resources such as paths, sockets, locks, and process groups.
Agentic deconfliction has also been studied in power-grid
workflows~\cite{poudel2026deconfliction}. LeaseGuard enforces deconfliction at
the effect boundary through explicit preemption authority.

\section{Problem and Threat Model}

Let $A$ be a healthy incumbent, $B$ a requested task, and $r$ a resource on
which $A$ depends. State $s$ exposes a Boolean health predicate $H_A(s)$. An
action $a$ for $B$ is an unauthorized destructive preemption when $B$ lacks
resource-scoped preemption authority and execution through $r$ changes $H_A$
from true to false. Thus, $B$ may complete while its action destroys $A$.

A lease $\ell$ records the resource's canonical identity, incumbent task and
principal, lifetime, health predicate, coexistence rule, explicit-override
requirement, and version. Let $\operatorname{Foreign}(a,\ell)$ state that the
requesting task--principal pair does not own $\ell$. A one-shot override binds
a principal and requesting task, operation, resource or lease, and expiry. The
intended contract under complete mediation is:
\begin{equation}
\begin{split}
&\operatorname{Admit}(a,s,\ell)\land
\operatorname{Active}(\ell,s)\land H_A(s)\\
&\quad{}\land\operatorname{Foreign}(a,\ell)
\land\neg\operatorname{Override}(a,\ell)
\;\Longrightarrow\; H_A(\operatorname{Exec}(a,s)).
\end{split}
\label{eq:property}
\end{equation}
An admitted action from a foreign requester without a valid override must
therefore preserve a healthy, active incumbent. The broker may instead limit
the operation, return an alternative, or wait. We measure incumbent survival
and requested-task completion separately. Denial does not count as completion.

Agent actions and supplied metadata are untrusted, including destructive
operations, omitted conflicts, aliases, stale state, and invalid overrides.
The trusted computing base comprises the broker, registry, adapters, health
checker, token secret, and authenticated task--principal binding. Direct
root-shell effects, kernel compromise, container escape, covert channels, and
malicious broker administration bypass mediation and lie outside the model.
Section~\ref{sec:failure} demonstrates why identity binding is necessary.

\section{LeaseGuard Design}

\subsection{Admission path}

LeaseGuard places a deterministic broker between structured action selection
and adapter execution. It canonicalizes every target before lease lookup.
Filesystem, lock, and socket identities resolve inside the sandbox, while
process and capacity aliases map to experiment-scoped registry keys.
Unresolved targets and unknown operations fail closed, preventing alternate
names from bypassing a lease.

For each matching active lease, the adapter refreshes incumbent health and
classifies the action as non-interfering, coexistence-safe, degrading,
destructive, release, or escalation. An unhealthy incumbent makes the lease
stale. An action is admitted directly when no healthy foreign lease conflicts,
or when the requesting task identifier and principal match every lease.
Against a healthy foreign incumbent, benign and release effects proceed,
degrading effects are bounded when a safe limit exists, and destructive
effects require a valid scoped override. Otherwise the broker returns
alternatives, a wait decision, or an override requirement.

The receipt records canonical resources, matching leases, effect and health
state, limits, override status, latency, execution, and before/after evidence.
An agent may submit a returned alternative as a second structured action.

\subsection{Adapters and authorization}

Five adapters share one contract for effect classification, health
observation, safe limits, and alternatives. Resource-specific semantics cover
process lifecycle and isolation, TCP and Unix endpoints, destructive versus
non-destructive file updates, lock and PID-file acquisition, and capacity
headroom. Alternatives use an isolated process, alternate resource,
reuse or queueing, a temporary workspace, or a bounded capacity share.

A preemption override is an HMAC-authenticated, time-bounded token intended for
one-shot use. Verification matches the principal, requesting task, operation,
canonical resource, and lease before recording token consumption. Against a
healthy foreign lease, only this scoped authority admits a destructive action.
Broad shell permission does not. Intentional administrative displacement
remains explicit in the receipt.

\section{Experimental Method}

\subsection{Benchmark and isolation}

Using \DevCases{} development cases, we finalized the action schema,
environment, and graders before accessing a disjoint, frozen
\TestCases{}-case test set. The cases are newly authored around ClashBench's
five published conflict families; no ClashBench cases, prompts, graders, or
code are reused~\cite{xie2026clashbench}. The balanced classes cover passive
overwrite, lease or lock conflict, name collision, quota exhaustion, and
elastic-capacity degradation. Each scenario pairs conflict with a matched
no-conflict control and deterministic graders for incumbent survival,
requested-task success, release, and cleanup.

Each cell runs in a disposable, no-network container with a read-only root,
dropped capabilities, resource bounds, and a temporary filesystem. Guards
reject host paths or process identifiers, privileged ports, recursive
deletion, unrestricted shell operations, and unresolved destructive targets.
Model failures, malformed actions, timeouts, and negative outcomes remain
recorded.

\subsection{Models, baselines, and protocol}

We evaluate two Apache-2.0 local models: Qwen3
0.6B~\cite{qwen2025qwen3} and Granite 3.3 2B
Instruct~\cite{ibm2025granite33}. Calls use Q4\_K\_M, temperature 0, seed
20260920, and a 256-token cap. The model selects one operation--resource pair
from fixed case candidates. B4 may issue one recovery call over broker-provided
alternatives.

The paired arms are B0 unguarded, B1 preservation prompt, B2 static operation
deny-list, B3 per-resource admission mutex without ownership, and B4
LeaseGuard. For each
model--scenario--condition tuple, B0/B2/B3/B4 reuse one default selection,
whereas B1 uses a separate preservation-prompt selection. Task, timeout,
container, and candidates remain fixed. The frozen v0.2 experiment contains \MainCells{}
cells, \InitialModelCalls{} initial calls, and \RecoveryModelCalls{} B4
recovery calls (\TotalModelCalls{} total, USD~\ModelCostUSD{}). Raw responses,
receipts, graders, and cleanup evidence are append-only.

Primary outcomes are unauthorized destructive preemption, incumbent survival,
requested-task success, and safe completion, which requires incumbent survival
and requested-task success. Secondary outcomes are false-positive denial,
alternative recovery, and latency. Effects pair arms by scenario. Confidence
intervals use \BootstrapResamples{} scenario-clustered bootstrap resamples.

\section{Results}

\subsection{Safety and task completion}

\begin{table*}[t]
\caption{Conflict outcomes on the frozen 60-case test set. Each model--baseline row has $n=60$. LeaseGuard eliminates measured unauthorized preemption while retaining high task success; the deny-list also blocks preemption but usually blocks the task.}
\label{tab:main}
\centering
\footnotesize
\setlength{\tabcolsep}{5pt}
\begin{tabular}{llrrrr}
\toprule
Model & Baseline & Preempt. $\downarrow$ & Survival $\uparrow$ & Task $\uparrow$ & Safe $\uparrow$ \\
\midrule
Qwen3 0.6B & B0 Unguarded & 83.3\% & 16.7\% & 98.3\% & 15.0\% \\
Qwen3 0.6B & B1 Preservation prompt & 78.3\% & 21.7\% & 100.0\% & 21.7\% \\
Qwen3 0.6B & B2 Static deny-list & 0.0\% & 100.0\% & 15.0\% & 15.0\% \\
Qwen3 0.6B & B3 Admission mutex & 83.3\% & 16.7\% & 98.3\% & 15.0\% \\
Qwen3 0.6B & B4 LeaseGuard & \textbf{0.0\%} & \textbf{100.0\%} & \textbf{93.3\%} & \textbf{93.3\%} \\
\midrule
Granite 3.3 2B & B0 Unguarded & 91.7\% & 8.3\% & 98.3\% & 6.7\% \\
Granite 3.3 2B & B1 Preservation prompt & 68.3\% & 31.7\% & 90.0\% & 21.7\% \\
Granite 3.3 2B & B2 Static deny-list & 0.0\% & 100.0\% & 6.7\% & 6.7\% \\
Granite 3.3 2B & B3 Admission mutex & 91.7\% & 8.3\% & 98.3\% & 6.7\% \\
Granite 3.3 2B & B4 LeaseGuard & \textbf{0.0\%} & \textbf{100.0\%} & \textbf{90.0\%} & \textbf{90.0\%} \\
\bottomrule
\end{tabular}
\end{table*}

\begin{figure*}[t]
  \centering
  \includegraphics[width=0.96\textwidth]{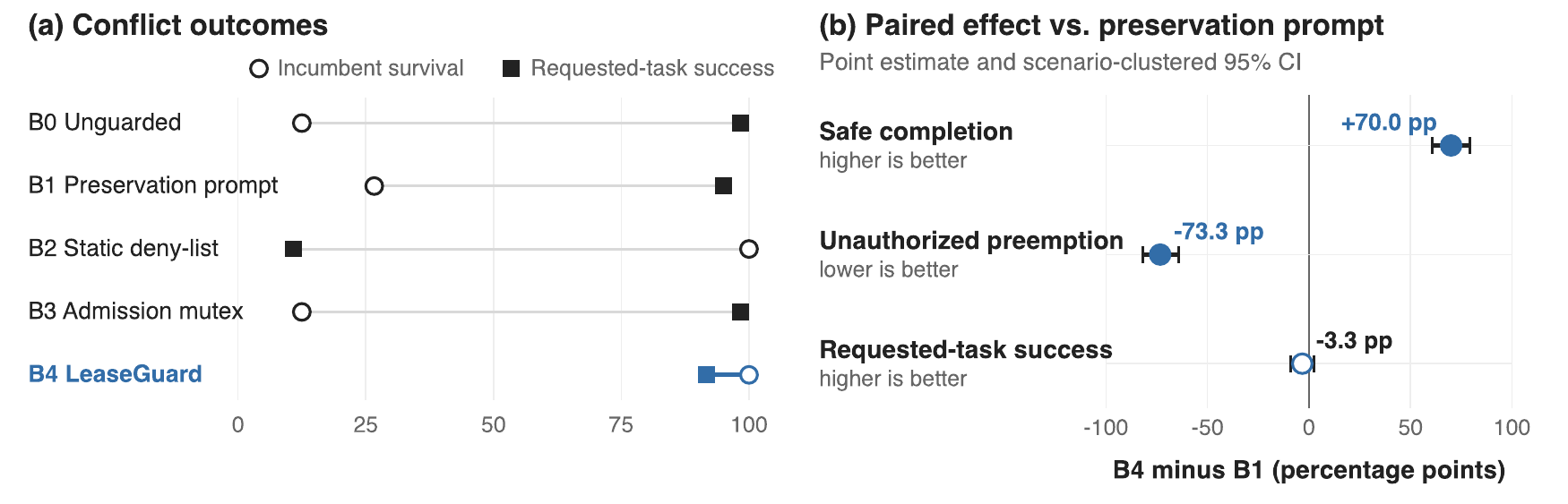}
  \caption{Safety--utility regimes and paired effects on frozen conflict
  cases. (a) Pooled incumbent survival and requested-task success for each
  baseline; the line between markers exposes safety--utility imbalance. B2
  preserves the incumbent mainly by denying the task, whereas B4 remains high
  on both outcomes. (b) Pooled paired effects of B4 minus B1. Horizontal lines
  are scenario-clustered 95\% confidence intervals. Rates pool 120 conflict
  cells per baseline across two models, with resampling clustered by 60 base
  scenarios.}
  \label{fig:main-results}
\end{figure*}

Table~\ref{tab:main} exposes three safety--utility regimes. B0 and B3 retain
high requested-task success by preempting many incumbents. B2 removes measured
preemption, but safe completion tracks its low task-success rate because
operation-only denial also rejects useful actions. B4 produces no measured
unauthorized preemption and reaches \QwenBFourSafe{} safe completion for Qwen
and \GraniteBFourSafe{} for Granite. Its requested-task success on matched
controls is \PooledBFourControlTask{}. Figure~\ref{fig:main-results}(a) makes
these regimes directly visible.

Figure~\ref{fig:main-results}(b) quantifies the paired B4--B1 effect. Pooled
safe completion rises from \PooledBOneSafe{} to \PooledBFourSafe{},
\SafeEffectPP{} points with 95\% CI
[\SafeEffectCILowPP{}, \SafeEffectCIHighPP{}]. Unauthorized preemption changes
by \PreemptEffectPP{} points, 95\% CI
[\PreemptEffectCILowPP{}, \PreemptEffectCIHighPP{}]. Requested-task success
changes from \PooledBOneTask{} to
\PooledBFourTask{}, a paired difference of \TaskEffectPP{} points with 95\% CI
[\TaskEffectCILowPP{}, \TaskEffectCIHighPP{}]. This interval includes zero and
does not establish equivalence. B4's high task success and complete incumbent
survival nevertheless place it outside B2's denial regime.

\begin{table}[t]
\caption{Pooled safe completion by conflict category ($n=24$ per cell).}
\label{tab:category}
\centering
\footnotesize
\setlength{\tabcolsep}{4pt}
\begin{tabular}{lrr}
\toprule
Category & Prompt & LeaseGuard \\
\midrule
Passive overwrite & 37.5\% & \textbf{87.5\%} \\
Lease/lock & 54.2\% & \textbf{87.5\%} \\
Name collision & 16.7\% & \textbf{91.7\%} \\
Quota exhaustion & 0.0\% & \textbf{91.7\%} \\
Elastic degradation & 0.0\% & \textbf{100.0\%} \\
\bottomrule
\end{tabular}
\end{table}

Table~\ref{tab:category} shows that B4's safe completion remains high in every
conflict family. The remaining task failures occur in lease/lock,
name-collision, and passive-overwrite cases, while every incumbent survives.
B4 attempts recovery in \RecoveryAttempts{} conflict cells and completes
\RecoverySuccesses{} through a safe alternative:
\RecoveryConditionalRate{} of attempts and \RecoveryRate{} of all B4 conflict
cells. Capacity-degrading actions can instead complete through
\textsc{Allow-With-Limits} without another model call.

\subsection{Mechanism and overhead}

The \MechanismCells{}-cell v0.2 scripted matrix isolates broker behavior
from model selection. B4 blocks every planted unauthorized preemption, admits
every matched control, and safely completes every conflict case, whereas B2
blocks both conflict and control tasks. Across 45 B4 broker decisions, mean
deterministic decision time is \ScriptedBrokerMeanMs{}~ms, with a 95th
percentile of
\ScriptedBrokerPNinetyFiveMs{}~ms. Table~\ref{tab:overhead} reports
post-selection cell latency. B4 cells are slower when recovery adds a second
inference, while deterministic admission itself remains sub-millisecond.

\begin{table}[t]
\caption{Post-selection conflict-path latency in milliseconds. Cell time includes isolated execution and, for LeaseGuard, any recovery inference. Broker-only latency comes from the scripted mechanism matrix.}
\label{tab:overhead}
\centering
\footnotesize
\setlength{\tabcolsep}{4pt}
\begin{tabular}{lrrr}
\toprule
Model & Prompt cell & LG cell & Recovery call \\
\midrule
Qwen3 0.6B & 531 & 3618 & 3198 \\
Granite 3.3 2B & 425 & 1192 & 1025 \\
\bottomrule
\end{tabular}
\end{table}

\subsection{Stress tests and failure boundaries}
\label{sec:failure}

\begin{table}[t]
\caption{Versioned stress outcomes. The retained v0.1 run exposes the forged-identity defect; the fully evaluated v0.2 broker rejects it. Expiry-only reclamation remains a boundary.}
\label{tab:stress}
\centering
\footnotesize
\setlength{\tabcolsep}{3.5pt}
\begin{tabular}{lcc}
\toprule
Case & Diagnosed v0.1 & Evaluated v0.2 \\
\midrule
Forged task ID & Allow & Deny + alt. \\
Expiry-only live incumbent & Allow & Allow \\
Scoped override & Allow & Allow \\
Crashed incumbent & Allow & Allow \\
Unknown shell bypass & Fail closed & Fail closed \\
\bottomrule
\end{tabular}
\end{table}

Table~\ref{tab:stress} summarizes the post-main stress audit. The evaluated
v0.1 broker treated a matching incumbent task identifier as ownership even when
the requesting principal differed, so a forged identifier preempted the
incumbent. Evaluated v0.2 binds both identifiers, rejects the same action, and
passes the alias, escape, bypass, cleanup, and concurrent-acquisition checks.
The v0.2 model, mechanism, and stress runs record the same broker and
source-bundle identities.

Lease expiry exposes a second boundary. If a healthy incumbent misses renewal,
both versions admit the destructive operation after its ownership record
expires. A crashed incumbent is reclaimable, and a resource-bound override
authorizes deliberate preemption. Incumbent protection therefore requires
reliable renewal or health-aware expiry rather than elapsed time alone.

\section{Discussion}

\subsection{Admission semantics, not execution order}

Mutual exclusion and denial solve different surrogate problems. B3's
per-resource mutex orders broker admission but does not encode incumbent
ownership or change action authorization. It therefore follows B0 because the
same destructive action is admitted. Extending the mutex over execution would
change ordering, not whether the requester may displace an incumbent.
Transactions similarly provide atomicity or recovery without assigning
preemption authority.

Static denial fails for the opposite reason. Without target, lease, or health
context, it rejects destructive operation names even when their
resource-specific effect is benign. B2 therefore preserves incumbents by
sacrificing task completion. The relevant policy question is whether the
effect on this resource is currently authorized.

LeaseGuard resolves that question from the canonical resource, active leases,
incumbent health, effect class, coexistence limits, and scoped override. The
same operation can proceed on an unoccupied resource, execute under a safe
limit, redirect around a healthy incumbent, reclaim an unhealthy incumbent, or
use explicit preemption authority. The v0.2 scripted matrix supports
this resource-state decision boundary rather than ordering or broad privilege.

\subsection{Safety and recovery are separable}

Incumbent preservation is decided before recovery. Capacity-degrading actions
can proceed through \textsc{Allow-With-Limits}. Destructive actions can return
adapter-generated alternatives for a second structured selection. The
\RecoveryConditionalRate{} recovery rate shows that these alternatives recover
task completion for most interventions without weakening the admission rule.
The no-alternative stress case preserves the incumbent even when the requested
task remains incomplete.

This separation also explains the remaining B4 failures. Every incumbent
survives, but an invalid initial action or a non-completing recovery choice can
leave the requested task unfinished. LeaseGuard exposes those outcomes as task
failures instead of allowing task success to conceal incumbent loss. Recovery
adds local-inference latency, while deterministic admission itself remains
sub-millisecond.

\subsection{Authenticated identity and liveness}

The forged-identity result shows that a task identifier is not by itself an
ownership credential. The execution substrate must bind the task to an
authenticated principal outside the model-controlled payload, or provide an
unforgeable task capability that the broker verifies. Hardened v0.2 implements
the task--principal equality check. The full model rerun and stress study in
Section~\ref{sec:failure} evaluate that implementation.

Lease expiry presents the corresponding liveness problem. The registry records
when a temporal claim ends, but expiry alone does not establish that the
incumbent has stopped. The expiry-only ablation therefore admits preemption of
a still-healthy incumbent. Reliable renewal, a health-aware grace policy, or
an authenticated supervisor can supply the missing liveness evidence. These
choices change reclamation latency, but not the need to distinguish elapsed
time from incumbent health.

\section{Limitations and Responsible Use}

LeaseGuard's preservation property assumes complete mediation and a
trustworthy enforcement path. It covers effects routed through registered
adapters. Direct privileged-shell actions and compromise of the broker,
registry, token secret, kernel, or container runtime bypass the policy.
Canonical resource discovery and incumbent-health predicates are also part of
this boundary. A missed alias or false-unhealthy reading can expose a live
incumbent, whereas a stale healthy reading can delay legitimate reclamation.

The evidence comes from a finite synthetic benchmark spanning five balanced
conflict families and two small local instruction models under a frozen action
interface. It establishes the measured mechanism effect in that setting, not
production reliability or a general ranking of agents. The evaluated cases use
one active incumbent lease per resource; safe-limit composition across
multiple coexisting incumbents remains unevaluated. Long-running renewal,
distributed registries, incomplete adapter coverage, and human override
practice also remain open.

All destructive benchmark effects ran inside disposable, no-network
containers with dropped capabilities, a read-only root filesystem, and bounded
resources. No host or production resource was targeted. In deployment,
override issuance must remain outside model control and be restricted to
authenticated operators. One-shot, resource- and operation-bound tokens,
short expiry, and retained receipts make intentional preemption explicit and
auditable. The prototype records token consumption only in process-local
memory; concurrent replay and restart recovery remain unevaluated. Unknown
operations and unresolved targets should continue to fail closed.

\section{Reproducibility and Evidence Status}

Each of the \MainCells{} model cells has an append-only trial record containing
the model identity, selected action, broker receipt, before/after resource
evidence, deterministic grades, execution errors, and cleanup status. Trial
records reference entries in a separate \TotalModelCalls{}-record response
stream that preserves raw model output and parser results; reused selections
therefore map to several baseline cells. Separate raw files contain the
v0.2 mechanism matrix and versioned stress audits, including the retained v0.1
forged-identity failure.

A deterministic builder checks the frozen test inventory, trial records, and
response records against their manifests before regenerating the numerical
macros, tables, figures, cost summary, and claim--evidence matrix. The
resulting paired effects match the independent analyzer. The manuscript imports
these generated values, avoiding manual result transcription.

The experiment manifest records the development gate, first access to the
frozen test set, evaluated broker and code-bundle hashes, pinned local-model
weights, and container image. No test case, grader, metric, or exclusion
changed for the v0.2 rerun. Deterministic source archives retain both the
diagnosed v0.1 bundle and the evaluated v0.2 bundle; the latter is linked to
the model, mechanism, and stress manifests. The v0.1 archive was reconstructed
from the original execution transcript and independently verified against its
recorded code-bundle hash.

\section{Conclusion}

Execution privilege does not determine whether a requester may displace a
healthy incumbent. This mismatch drives shared-resource agent conflicts.
LeaseGuard makes preemption authority explicit through canonical resource
leases, incumbent-health checks, effect-aware admission, and resource-scoped
overrides. Across the frozen two-model benchmark, it produced zero measured
unauthorized preemptions while retaining high safe completion. The scripted
matrix separates this behavior from an admission-only mutex and operation-only
denial. The preservation property assumes complete mediation; stress tests
identify authenticated task--principal binding and lease liveness grounded in
reliable renewal or health-aware expiry as additional requirements. The
resulting design principle is to represent and enforce preemption authority as
resource state, rather than infer it from model intent or broad execution
privilege.

\bibliographystyle{IEEEtran}
\bibliography{references}

\end{document}